\documentclass[preprint]{revtex4-1}

\usepackage{amsmath}
\usepackage{amsfonts}
\usepackage{amssymb}
\usepackage{graphicx}
\usepackage{hyperref}
\usepackage{mhchem}
\usepackage{nameref}
\usepackage{caption}
\usepackage{subcaption}
\usepackage{color}
\providecommand{\e}[1]{\ensuremath{\times 10^{#1}}}

	\newcommand{\ncd}{\newcommand}
	\ncd{\mrm}    {\mathrm}
	\ncd{\beq} {\begin{equation}}
	\ncd{\eeq} {\end{equation}}

	\def\d{{\rm d}}

\begin{document}

	\title{Geometric description of chemical reactions}

\author{Hernando Quevedo$^{1,2,3}$ and Diego Tapias$^{1,2}$}
		\email{quevedo@nucleares.unam.mx,diego.tapias@nucleares.unam.mx}
		\affiliation{
     $^1$Instituto de Cosmologia, Relatividade e Astrofisica ICRA - CBPF\\
          Rua Dr. Xavier Sigaud, 150, CEP 22290-180, Rio de Janeiro, Brazil\\
		$^2$Instituto de Ciencias Nucleares, Universidad Nacional Aut\'onoma de M\'exico, AP 70543, M\'exico DF 04510, Mexico \\
		$^3$Dipartimento di Fisica and ICRA, Universit\`a di Roma La Sapienza, Piazzale Aldo Moro 5, I-00185 Rome, Italy
}

	\begin{abstract}

We use the formalism of Geometrothermodynamics  to describe chemical reactions in the context of equilibrium thermodynamics.
Any chemical reaction in a closed system is shown to be described by a geodesic in a $2-$dimensional manifold that  can be
interpreted as the equilibrium space of the reaction. We first show this in the particular cases of a reaction with only two species corresponding to either two ideal gases or two van der Waals gases. We then consider the case of a reaction with an 
arbitrary number of species. 
The initial equilibrium state of the geodesic is determined by the initial
conditions of the reaction. The final equilibrium state, which follows 
from a thermodynamic analysis of the reaction, is shown 
to correspond to a coordinate singularity of the thermodynamic metric 
which describes the equilibrium manifold.

	\end{abstract}

\maketitle


\section{Introduction}

The geometric nature of the thermodynamics has been considered as an important question since the pioneering works of Gibbs \cite{gibbs} and Caratheodory \cite{cara}. However, it  was the development of differential geometry and Einstein's theory of General Relativity which increased the interest to extend and develop the geometric approach to other branches of physics. 
Particularly, in thermodynamics, the work of Hermann and Mrugala \cite{her73,mrugala1,mrugala2} set up the manifold called \emph{thermodynamic phase space}, 
 where   the ``contact geometry'' approached by Gibbs and Caratheodory becomes meaningful. 
Riemannian geometry was first introduced in the {\it space of equilibrium states} by Rao \cite{rao45}, in 1945, 
by means of a metric whose components in local coordinates coincide with Fisher's
information matrix. In fact, the Fisher-Rao metric can be considered as an element of the class of so-called Hessian metrics whose local components coincide with the Hessian of any arbitrarily chosen thermodynamic potential. 
Rao's original work has been followed up and extended by a number of
authors (see, e.g., \cite{amari85} for a review).
The proposal coined by Quevedo \cite{quev07} as Geometrothermodynamics (GTD) was essential to unify both approaches and to endow the equilibrium states manifold with a Legendre invariant metric. 

The importance of Legendre invariance lies in the thermodynamics itself, meaning that once a representation is chosen to describe the system (for instance, the internal energy or entropy), its Legendre transform (e.g., the Gibbs free energy or the Massieu-Planck potential)  contains the same information as the original representation.
Therefore,  Legendre invariance should be an essential ingredient of a geometric construction.
 
Basically, in the GTD approach, the thermodynamic phase space is endowed with a  Legendre invariant metric, and its maximally integral submanifold, that inherits its metric structure, is identified with the space of equilibrium states. In essence, a point of this space corresponds to an equilibrium state and, therefore, the thermodynamic processes take place in the equilibrium manifold. Consequently, one expects the geometric properties of the equilibrium manifold to be related to the macroscopic physical properties. The details of this relation can be summarized in three points:
\begin{itemize}
\item The curvature of the equilibrium  manifold reflects the thermodynamical interaction.
\item The phase transitions correspond to curvature singularities.
\item There exists a correspondence between quasi-static thermodynamic processes and certain geodesics of the equilibrium manifold.
\end{itemize}

The geometrothermodynamic approach has been applied to classical systems such as the ideal gas \cite{quev07} and the van der Waals gas \cite{qr12}, to more exotic systems such as black holes \cite{qstv10a}, and also in the context of relativistic cosmology to describe the evolution of our Universe \cite{abcq12}. In all the cases, in which the analysis have been performed completely, the summarized items have been confirmed and the results have been shown to be Legendre invariant.   

In this paper, we aim to describe the geometry behind a chemical reaction. First, we will consider the case of a reaction with only two species, and then we will show that results can be generalized to include any arbitrary finite  number of species.  
 The paper is organized as follows. In section \ref{theory}, we present a review of the geometrothermodynamic structures and the classical thermodynamics approach to chemical reactions; particularly, we analyze the reaction \ce{A_{(g)}  <=> B_{(g)}} in the context of classical thermodynamics. 
 In sections \ref{sec:ig}  and \ref{sec:vdw}, we consider the case of a reaction with two species corresponding to ideal gases and 
 van der Waals gases, respectively. In both cases, we present the thermodynamic analysis of the reaction to find the final state of equilibrium, and analyze the same situation from the point of view of GTD. It is shown that each reaction can be represented by a geodesic in the equilibrium manifold. Then, in section \ref{sec:gtdcr}, we show the applicability of  GTD to a general chemical reaction. Finally, section \ref{sec:con}
 is devoted to the conclusions.

\section{The theory}
\label{theory}

\subsection{Geometrothermodynamics}

The idea behind the geometrization of a thermodynamic system is simple: to build a space where each point corresponds to an equilibrium state. The physics behind the equilibrium thermodynamics allows us to say that this space is an $n$-dimensional manifold with the dimension corresponding to the number of thermodynamic degrees of freedom of the system. In consequence, one needs only $n$ independent variables  to coordinatize the manifold. 

In standard equilibrium thermodynamics, to a system with $n$ degrees of freedom it is possible to associate $n$ extensive variables $E^a$, $n$ intensive variables $I^a$, where the index $a$ runs from 1 to $n$, and a thermodynamic potential $\Phi$, relating them. In this context, the terms extensive and intensive are general concepts that refer to the independence or dependency of the variables associated to a given potential. For example, for a closed simple system with two degrees of freedom, the independent variables are $T$ and $P$, if the potential chosen to describe the system is $G$, or  $U$ and $V$, if the fundamental potential is $S$. Recall that $S=S(U,V)$.

Consequently, from the point of view of thermodynamics, to a system with $n$ degrees of freedom  we associate $2n+1$ variables, $n$ of them being independent. Geometrically, this idea corresponds to an embedding $\varphi$ of an $n$-dimensional manifold $\mathcal{E}$ into a $(2n+1)$-dimensional manifold $\mathcal{T}$ given by
\beq
	\label{gtd.emb}
	\varphi:\mathcal{E}\longrightarrow\mathcal{T},
	\eeq
or, in coordinates,
\beq
\varphi:\{E^a\} \longrightarrow \{\Phi(E^a),I^b(E^a),E^a\}\ ,
\eeq
where $b$ also goes from $1$ to $n$.
The manifold $\mathcal{T}$ is  a contact manifold \cite{her73}. This means that ${\cal T}$ is endowed with a family of tangent hyperplanes 
(contact structure) defined by the so-called fundamental 1-form $\Theta$ that satisfies the non-integrability condition
	\beq
	\label{gtd.ni}
	\Theta \wedge (\d \Theta)^n \neq 0\ .
	\eeq
The manifold $\mathcal{T}$ is called the \emph{thermoynamic phase space}. It can be coordinatized by the $2n+1$ variables $\{Z^A\}=\{\Phi,E^a,I^a\}$, where $A=0,...,2n$. Its existence is relevant because it results adequate to perform  Legendre transformations (as in thermodynamics) as a change of coordinates. Formally, a Legendre transformation is a contact transformation, i.e., a transformation which leaves the contact structure unchanged; in coordinates $\{Z^A\}$, it is defined as \cite{arnold}
\begin{eqnarray}
\{Z^A\} \rightarrow \{\tilde{Z}^{A}\} = \{\tilde{\Phi},\tilde{E}^a,\tilde{I}^a\} 
\end{eqnarray}
\beq
\Phi = \tilde{\Phi} - \delta_{kl} \tilde{E}^k\tilde{I}^l \, , \quad E^i = -\tilde{I}^i \, , E^j = \tilde{E}^j \, ,\  I^i = \tilde{E}^i \, , \ I^j = \tilde{I}^j\ ,
\eeq
where $i \cup j$ is any disjoint decomposition of the set of indices $\{1, \ldots, n\}$, and $k,l = 1, \ldots, i$.

According to the Darboux theorem \cite{her73}, the 1-form $\Theta$ of equation \eqref{gtd.ni} can be given in the coordinates $\{Z^A\}$ as:
\beq
	\label{gtd.local}
	\Theta = \d \Phi - I_a \d E^a 
	\eeq 
where we use Einstein's summation convention for repeated indices. It can easily be seen that after a Legendre transformation, the new 1-form $\tilde{\Theta}$ in coordinates $\{\tilde{Z}^A\}$ reads
\beq
	\label{gtdlegendre}
	\tilde{\Theta} = \d \tilde{\Phi} - \tilde{I}_a \d \tilde{E}^a \ .
	\eeq
This proves that the contact structure remains unchanged.

On other hand, the manifold $\mathcal{E}$ is the maximally integral submanifold of $\mathcal{T}$, and is defined in such a way that the properties of the thermodynamic systems are encoded in it. So far, we have seen that $\mathcal{E}$ is specified through the embedding \eqref{gtd.emb}, which is equivalent to specifying the fundamental equation of the system $\Phi(E^a)$. The next step is to introduce the relations of standard equilibrium thermodynamics into the definition of the manifold. To this end, we demand that the embedding \eqref{gtd.emb} satisfies the condition
	\beq
	\label{gtd.01}
	\varphi^*(\Theta) = 0\ ,
	\eeq
where $\varphi^*$ is the pullback of $\varphi$. In coordinates, it takes the form
	\beq
	\label{gtd.02}
	\varphi^*(\Theta) = \varphi^*\left(\d \Phi - I_a \d E^a\right) = \left(\frac{\partial \Phi}{\partial E^a} - I_a\right)\d E^a  = 0.
	\eeq
It follows immediately that
	\beq
	\label{gtd.03}
	\Phi = \Phi(E^a) \quad \text{and} \quad \frac{\partial \Phi}{\partial E^a} = I_a\ .
	\eeq
Equations \eqref{gtd.02} and \eqref{gtd.03} constitute the standard Gibbs relations of equilibrium thermodynamics in $\mathcal{E}$, namely,
	\beq
	\d \Phi = I_a \d E^a \ .
	\eeq
The manifold $\mathcal{E}$ defined in this way is called the \emph{space of equilibrium states}.

In addition to the geometric description of thermodynamics in terms of a contact structure, the GTD program promotes the contact manifold $(\mathcal{T},\Theta)$ into a Riemannian contact manifold $(\mathcal{T},\Theta,G)$, where $G$ is a metric sharing the symmetries of $\Theta$. The most general metric invariant under total and partial Legendre transformations that has been found so far is given by \cite{vqs10}
	\beq
	\label{quevedoGIII}
	G  =\Theta \otimes \Theta +\Lambda\left(Z^A\right) \sum_{a=1}^n \left[\left(E^a I_a \right)^{2k+1} \d E^a \otimes \d I_a \right],
	\eeq
where $\Lambda(Z^A)$ is an arbitrary Legendre invariant function of the coordinates $Z^A$ and $k$ is an integer.
The corresponding induced metric in the space of equilibrium states is given by
	\begin{equation}
	\label{quevedogII}
	g_{\Phi}=\varphi^*(G) = \Lambda(E^a) \sum_{a,b=1}^n\left[\left(E^a \frac{\partial \Phi}{\partial E^a} \right)^{2 k + 1}  \frac{\partial^2 \Phi}{\partial E^a \partial E^b}\ \d E^a \otimes \d E^b\right]\ .
	\end{equation} 
As suggested in \cite{vqs10}, this metric could be useful to analyze multicomponent systems, particularly systems where  chemical reactions take place. We will show in the next sections that, in fact,  chemical reactions can be represented as geodesics of the equilibrium manifold described by the metric (\ref{quevedogII}).  

Notice that to compute the explicit components of this metric, it is necessary to specify only the fundamental equation $\Phi=\Phi(E^a)$. 
Thus, all the geometric properties of the equilibrium space are determined by the fundamental equation. This is similar to the situation in classical thermodynamics where the fundamental equation is used to determine all the equations of state and thermodynamic properties of the system.

Notice that the metric $G$ contains the arbitrary parameter $k$ which, however, can be absorbed by renaming the coordinates as $dX^a = (E^a)^{2k+1} d E^a$ and 
$dY_a = (I_a)^{2k+1} d I_ a$. Then,
	\beq
	\label{GIIIXY}
	G  =\Theta \otimes \Theta +\Lambda(Z^A) \sum_{a=1}^n \left( \d X^a \otimes \d Y_a \right) \ .
	\eeq
Furthermore, the arbitrary function $\Lambda(Z^A)$ can be fixed by demanding invariance with respect to changes of representation, an issue which is outside of the scope of the present work \cite{blnq13}. It is therefore possible to perform the entire analysis with an arbitrary function $\Lambda(Z^A)$ in coordinates $Z^A=(\Phi,X^a, Y_a)$; however, the physical interpretation of these coordinates becomes cumbersome and makes it difficult the physical interpretation of the results. For the sake of simplicity, we will use in this work the 
particular choice $k=-1$ and $\Lambda=-1$, which has been  shown to be  useful also to describe geometrically systems like the ideal gas or van der Waals gas \cite{vqs10}. Then, in the particular case $n=2$, the metric of the equilibrium manifold reduces to 
\begin{eqnarray}
g_\Phi &=& - \left(E^1 \frac{\partial \Phi}{\partial E^1}\right) ^{-1} \frac{\partial^2 \Phi}{\partial (E^1)^2} d E^1 \otimes d E^1 - 
\left(E^2 \frac{\partial \Phi}{\partial E^2}\right) ^{-1} \frac{\partial^2 \Phi}{\partial (E^2)^2} d E^2  \otimes dE^2  \notag \\
& -&  \left[ \left(E^1 \frac{\partial \Phi}{\partial E^1}\right) ^{-1} + \left( E^2 \frac{\partial \Phi}{\partial E^2}\right) ^{-1} \right] \frac{\partial^2 \Phi}{\partial E^1 \partial E^2} d E^1  \otimes d E^2 \ .
\label{metrica}
\end{eqnarray}

To analyze the geometric properties of the equilibrium manifold, we will consider the connection and the curvature. In particular, the connection is used to represent the geodesic equations in a given coordinate system
\beq
\frac{d^2E^a}{d\tau^2} + \Gamma^a_{\ bc} \frac{dE^b}{d\tau} \frac{dE^c}{d\tau} = 0 \ , \quad
\Gamma^a_{\ bc} = \frac{1}{2}g^{ad}\left(\frac{\partial g_{db}}{\partial E^c} + \frac{\partial g_{dc}}{\partial E^b} -
\frac{\partial g_{bc}}{\partial E^d}\right) \,
\eeq
where $\Gamma^a_{\ bc}$ are the Christoffel symbols. The solutions of these equations are the geodesics $E^a(\tau)$, where $\tau$ is an
affine parameter along the trajectory. One of the main goals of the present work is to show that a given chemical reaction can be
represented geometrically as a family of geodesics of the equilibrium manifold, which is 
determined by the fundamental equation of the chemical system. 

The curvature tensor is defined as 
\beq 
\label{curvature}
R^a_{\ bcd} = \frac{\partial \Gamma^a_{\ bd}}{\partial x^c} - \frac{\partial \Gamma^a_{\ bc}}{\partial x^d} +
\Gamma^a_{\ e c}\Gamma^e_{\ bd} - \Gamma^a_{\ e d}\Gamma^e_{\ bc} \ .
\eeq
In GTD, the curvature tensor of the equilibrium manifold is expected to be
a measure of the interaction between the components of the thermodynamic system. 
Furthermore, from the curvature tensor one can define the Ricci tensor $R_{ab} = g^{cd} R_{acbd}$
and the curvature scalar $R = g^{ab}R_{ab}$. Notice that in the case of a two-dimensional space,
the curvature tensor has only one independent component, say $R_{1212}$ and, therefore, the Ricci tensor
and the curvature scalar are proportional to $R_{1212}$.

\subsection{Thermodynamics of chemical reactions}
\label{resthermo}

Consider the general chemical reaction
\beq
 \ce{a_1A_1 + a_2A_2 + $\ldots$ <=> b_1B_1 + b_2B_2 + $\ldots$}\ ,
\label{cr}
\eeq
in which the $a_1,\ a_2, ...$ are the stoichiometric numbers of the reactants $A_1,\ A_2,...$, and $b_1,\ b_2,...$ are the stoichiometric coefficients of the products $B_1,\ B_2, ...$, respectively. Notice that the species need not all to occur in the same phase. The main condition for the chemical-reaction equilibrium in a closed system  is that  \cite{Levine}
\beq
\label{equilib}
\sum_i \nu_i\mu_i = 0 \ ,
\eeq
where the coefficients $\nu_i=(-a_1,-a_2,...,b_1,b_2,...)$ refer to the stoichiometric numbers and $\mu_i=(\mu_{A_1},...,\mu_{B_1},...)$ represents the chemical potential of the $i-$species. Notice that the condition (\ref{equilib}) holds no matter how the closed system reaches its final equilibrium state. 
During a reaction, the change of numbers of moles of the species $i$, $\Delta n_i = n_i - n_{i,0}$, where $n_{i,0}$ is the 
number of moles of species $i$ at the beginning of the reaction, is proportional to the stoichiometric number $\nu_i$
 with the extent of reaction $\xi$ as the proportionality factor, $\Delta n_i = \nu_i \xi$. For an infinitesimal extent 
 of reaction $d\xi$, it holds that $dn_i = \nu_i d\xi$. The fact that the extent of reaction can be treated as an infinitesimal quantity is essential for the geometric description we will present below. 

The main premise to apply classical thermodynamics in closed systems, where chemical reactions can occur, is that we can use thermodynamic variables to describe the system even if it is not in material equilibrium. It means that variables such as $U$, $S$, $T$, $V$, etc., are completely defined at any extent of reaction.

Though \eqref{equilib} is useful for practical situations, it does not contain information about the behavior of the different thermodynamic potentials from initial conditions to equilibrium. To obtain this information, we will use the fundamental equation of the chemical system. Let $\Phi_{A_j}(E^a_{A_j})$ represents the fundamental equation of the species $A_j$. Then, the fundamental equation for the general reaction (\ref{cr}) can be constructed as follows 
\begin{eqnarray}
\label{total}
\Phi(E^a_{A_1},E^a_{A_2}, \ldots) &=& \Phi_{A_1} (E^a_{A_1}) + \ldots + \Phi_{B_1}(E^a_{B_1}) + \ldots   +\Phi_{{A_1},{A_2}}(E^a_{A_1},E^a_{A_2})+\ldots  \notag \\
&+& \Phi_{{A_1},{B_1}}(E^a_{A_1},E^a_{B_1}) + \ldots + \Phi_{{A_1},{A_2},{B_1}}(E^a_{A_1},E^a_{A_2},E^a_{B_1}) + \ldots
\end{eqnarray}
Notice that in this expression we are taking into account all possible interactions between all the species. The only assumption is that the interaction between the species $A_j$ and $A_k$ depends on the variables $E^a_{A_j}$ and $E^a_{A_k}$, only. It seems that this condition is not very restrictive in realistic situations.

In the last subsection, we emphasized  the role of the Legendre transformations from the geometrical point of view. Now, we will mention their importance from the thermodynamic point of view.
\begin{enumerate}
\item It does not matter which potential $\Phi(E^a)$ is chosen to describe a particular system, all of them will contain the same thermodynamic information. 
\item The prediction of the final equilibrium state is made in accordance with the ``extremum principles"; these principles could be different for different potentials. The importance of the Legendre transformations is that they always can be used to find a potential in which the extremum principles hold for the experimental working conditions.
\end{enumerate}
For later use, we summarize here the ``extremum principles'' \cite{Callen}: 
\begin{itemize}
\item \emph{Entropy Maximum Principle.} The equilibrium value of any internal unconstrained parameter is such as to maximize the entropy for the given value of the total internal energy.
\item \emph{Energy Minimum Principle.}  The equilibrium value of any internal unconstrained parameter is such as to minimize the energy for the given value of the total entropy.
\item \emph{General Minimum Principle for the Legendre Transforms of the Energy.} The equilibrium value of any unconstrained internal parameter is such as to minimize the Legendre transform of the internal energy for a constant value of the transformed variable(s).
\item \emph{General Maximum Principle for the Legendre Transforms of the Entropy.} The equilibrium value of any unconstrained internal parameter is such as to maximize the Legendre transform of the entropy for a constant value of the transformed variable(s).
\end{itemize}
Consequently, we can talk about Legendre invariance in two senses. Firstly, in the sense that the thermodynamic information remains conserved under a Legendre transformation and, secondly, 
in the sense that the same equilibrium value for one or several internal unconstrained parameters will be obtained, independently of the thermodynamic potential $\Phi$, under the condition that the experimental restrictions are according to the particular restrictions contained in the ``extremum principle'' for $\Phi$.

In the following sections, we will study the chemical reaction \ce{A_{(g)} <=> B_{(g)}} considering $A$ and $B$ either as ideal monoatomic gases  or as van der Waals monoatomic gases.

\section{Ideal gases}
\label{sec:ig}

\subsection{Thermodynamics}

If we consider the species as ideal gases, 
the corresponding fundamental equation for each species in the entropy representation reads \cite{Callen}:
\beq                
\label{ecfunds}
S(U,V,n) = n s_0 + nR \ln\left[\left(\frac{U}{U_0}\right)^c \left(\frac{V}{V_0}\right) \left(\frac{n}{n_0}\right)^{-(c+1)} \right]\ ,
\eeq
where $s_0$, $U_0$, $V_0$ and $n_0$ refer to the standard values of reference, $c$ is a dimensionless constant related to the heat capacity of the ideal gas, i.e., $C_{v,n}=c\,R$, and $R$ is the universal gas constant. 

\begin{figure}
        \centering
     \begin{subfigure}[b]{0.3\textwidth}
                \centering
               \includegraphics[scale=0.3]{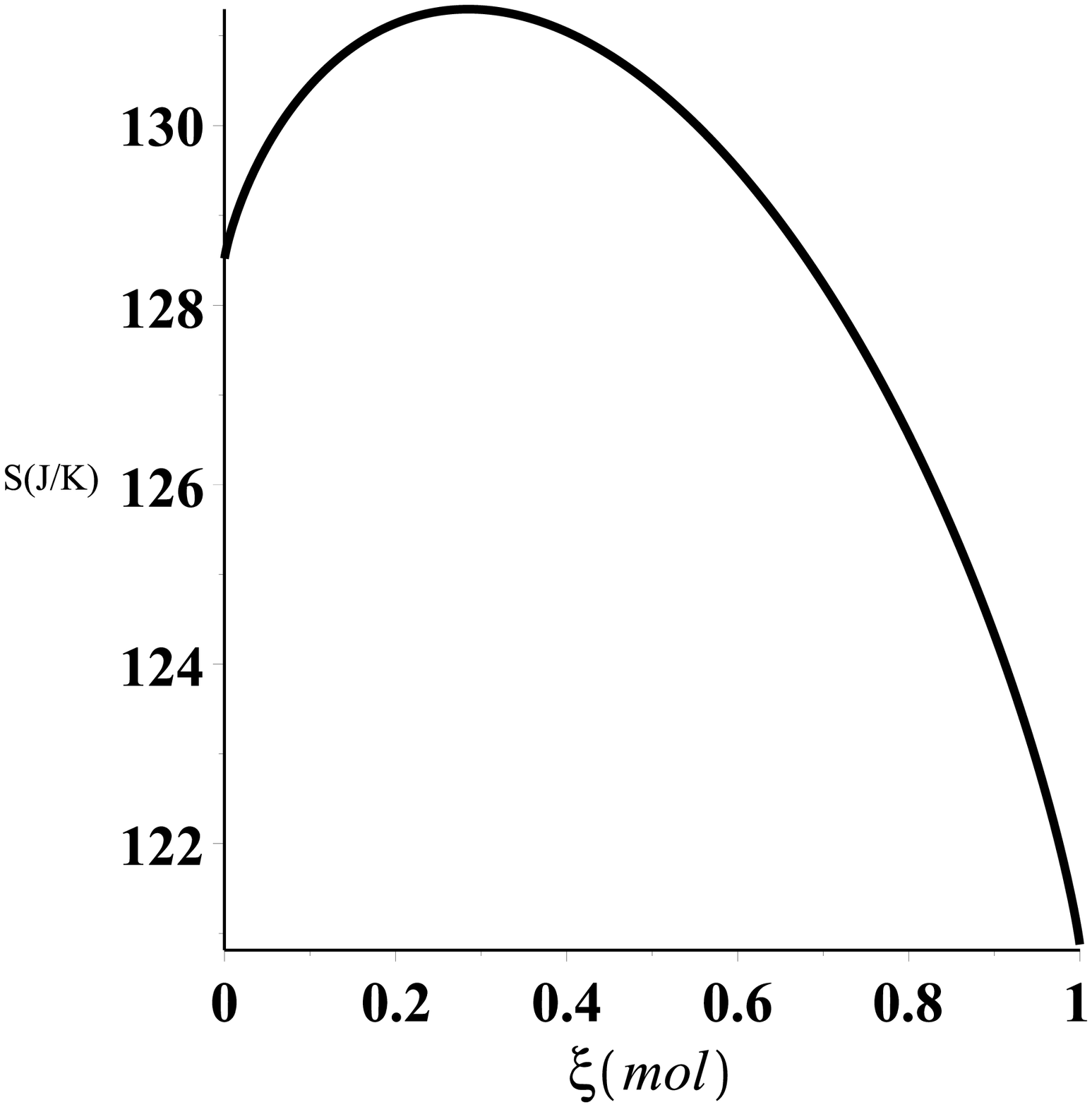}
\caption{Evolution of the Entropy potential}
\label{compa1}
        \end{subfigure}
\qquad
        \begin{subfigure}[b]{0.3\textwidth}
               \centering
\includegraphics[scale=0.3]{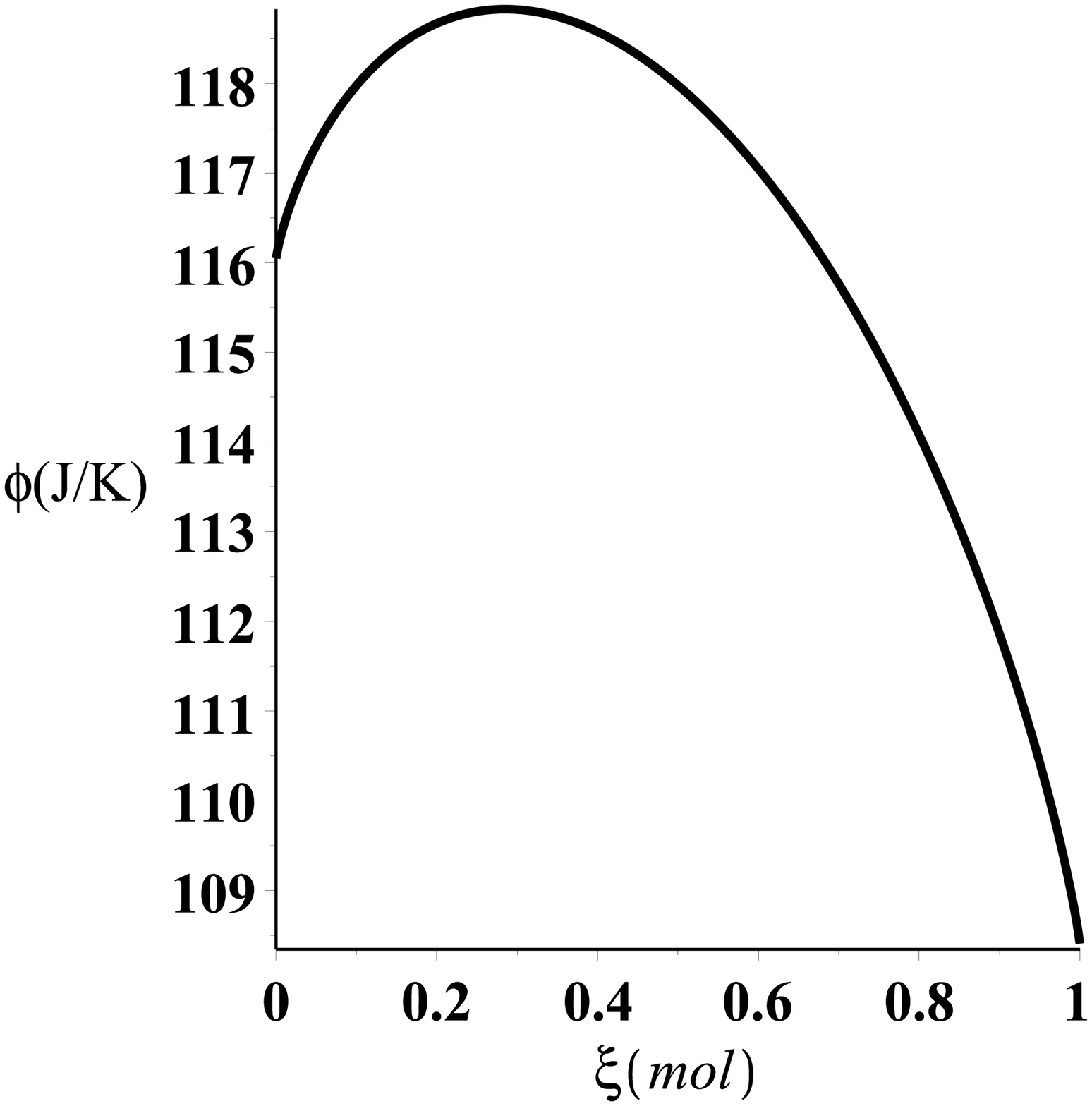}
\caption{Evolution of the Massieu potential}
\label{compa2}
        \end{subfigure}
%
%
        ~ 
%
             \caption{Behaviour of the thermodynamic potentials for the reaction \ce{A_{(g)} <=> B_{(g)}} at T=300 K considering \ce{A} and \ce{B}  as ideal gases. 
             }\label{comparación}
\end{figure}

Let us consider the particular case in which only two species take part in the reaction.
It turns out that it is convenient to study the evolution of the total entropy in terms of the extent of reaction. 
According to Eq.\eqref{total},  the fundamental equation takes the form
\begin{eqnarray}
&S&(U_A,V_A,U_B,V_B,\xi) = S(U_A,V_A,\xi) + S(U_B,V_B,\xi) \notag \\
&=&  (n_{A,0}-\xi) s_{0,A} + (n_{A,0}-\xi) R \ln\left[\left(\frac{U_A}{U_{0,A}}\right)^{c_A} \left(\frac{V_A}{V_{0,A}}\right) \left(\frac{n_{A,0}-\xi}{n_{0,A}}\right)^{-(c_A+1)} \right] \notag  \\  &+& (n_{B,0}+\xi) s_{0,B} + (n_{B,0}+\xi) R \ln\left[\left(\frac{U_B}{U_{0,B}}\right)^{c_B}\left(\frac{V_B}{V_{0,B}}\right) \left(\frac{n_{B,0}+\xi}{n_{0,B}}\right)^{-(c_B+1)} \right] \ ,
\label{fundamental}
\end{eqnarray}
where we neglected the interaction term for simplicity, and introduced explicitly the variable extent of reaction $\xi$.
Note that $n_{i,0}$ refers to the initial conditions and $n_{0,i}$ refers to the values of the state of reference.
For simplicity the values for $n_{0,i}$ and $V_{0,i}$  are set equal to one. 
Moreover, the values for $U_{0,i}$ and $s_{0,i}$ -which depend on the nature of each gas \cite{Tschoegl}- are chosen in such a way that they basically take into account the differences between the species. These and other values are shown in Table \ref{condiciones}.
\begin{table}[h]
\begin{center}
\begin{tabular}{|l | c | c|}
\hline
Gas & \ \ A \ \ & \ \ B \ \ \\
\hline
Initial moles (mol) & 1& 0 \\
\hline
Heat capacity $c$ & $\frac{3}{2}$ & $\frac{3}{2}$ \\
\hline
Molar standard entropy (J/mol-K) & 1 & 2 \\
\hline
Standard internal energy (J) & 1 & 2  \\
\hline
\end{tabular}
\caption{Conditions for the gases A and B}
\label{condiciones}
\end{center}
\end{table}
Finally, we take the temperature as $T_A=T_B=T=300 K$ and the volume as $V_A=V_B=V=20 L$, experimental conditions that can easily be achieved. With these conditions, the fundamental equation reduces to
\beq
S(\xi) = 1+\xi+ R \, \left( 1-\xi \right) \ln  \left(\frac{4.58\e{6}}{
1-\xi} \right) + R \,\xi\,\ln  \left({
\frac {1.62\e6}{\xi}} \right) \ ,
\label{trabajo}
\eeq
where the value of the constants have been rounded to simplify the presentation. 
A plot of this function is displayed in figure \ref{compa1}. 

Since the temperature is constant and $c_A= c_B$, the total internal energy is constant. Thus, according to the ``entropy maximum principle", we have  that the maximum of $S$, as a function of the extent of reaction, corresponds to the equilibrium condition. 
In this case, $S$ reaches it maximum value $S_f$ at $\xi=\xi_f\approx 0.285$.

To illustrate the significance of Legendre invariance, we can analyze other potentials that are obtained from 
$S$ by means of Legendre transformations. 
Consider, for instance, the Massieu potential (Helmholtz free energy) 
\begin{eqnarray}
\phi &:=& S -\frac{1}{T}U   \notag \\ &=& \frac{p}{T}V -\sum_i \frac{\mu_i}{T}n_i 
\end{eqnarray}
Considering the assumption \eqref{total} with no interacting term, we obtain
\begin{eqnarray}
\phi(\beta,V,n_A,n_B) &=& \phi_A(\beta,V,n_A)+  \phi_B(\beta,V,n_B)  \notag \\ &=& \frac{p_A}{T}V + \frac{p_B}{T}V - \frac{\mu_A}{T}n_A - \frac{\mu_B}{T}n_B
\end{eqnarray}
where $\beta =\dfrac{1}{T}$. To calculate explicitly this function we use
\begin{eqnarray}
\mu_i(U_i,V,n_i)&=& -T\frac{\partial S_i}{\partial n_i} \\
p_i(U_i,V,n_i)&=&T\frac{\partial S_i}{\partial V} \\
U_i=\dfrac{c_iRn_i}{\beta}
\end{eqnarray}
In this way, the fundamental equation of the system is:
\begin{eqnarray}
\phi\left(\beta,V, n_A, n_B \right)= \sum_{i =A,B}  n_i\left\{s_{0,i}+ R\ln  \left[  \left( {\frac {\beta_0^i}{{ \beta}}} \right) 
^{c} \left( {\frac {n_i}{{ n_{0,i}}}} \right) ^{-1}\left(\frac{V}{{ V_0^i}}\right)
 \right] - cR  \right\}\ .
\label{massig}
\end{eqnarray}

Taking into account the considerations discussed for the entropy representation, and the value $\beta = \dfrac{1}{300} K^{-1}$, finally the fundamental equation in this representation is reduced to
\beq
\phi (\xi) = -11.47 +\xi+ R\, \left( 1-\xi \right) \ln  \left( 
{\frac {4.58 \e 6}{1-\xi}} \right) + R\,\xi\,\ln 
 \left( {\frac {1.62 \e 6}{\xi}} \right)\ ,
 \eeq
where again we have rounded the values of the constants for simplicity. The plot of this function is displayed in figure \ref{compa2}.
In this case, the transformed variable of the Legendre Transform is $\beta$, and since it is a constant, the maximum of $\phi (\xi)$ corresponds to the equilibrium state, which is reached at $\xi=\xi_f\approx 0.285$. Thus, in agreement with the Legendre invariance, we obtain the 
same result as in the entropy representation.

The main result of this thermodynamic analysis is that a system with initial parameters as given in Table \ref{condiciones}, which correspond to an entropy $S_{in}$,  undergoes a chemical reaction whose final state is the equilibrium state characterized by the reaction extent $\xi\approx 0.285$ and the entropy $S_f$. Notice that  for $\xi_{in}<\xi_f$ ( $\xi_{in}>\xi_f$)  all the states with $\xi>\xi_f$ ($\xi<\xi_f$)  are not permitted. In fact, once the system reaches the final equilibrium state at $\xi=\xi_f$, the chemical reaction ends, and the states characterized by  a decrease of entropy are unphysical   according to the second law of the thermodynamics.

\subsection{Geometrothermodynamics}

Recall that to construct the metric of the equilibrium manifold we only need the fundamental equation. 
Under the restrictions corresponding to the chemical reaction of two ideal gases as described in the last subsection, 
the original fundamental equation  (\ref{fundamental})
reduces to a function that depends on two variables only, namely,
\begin{equation}
S(U,\xi) = 1+\xi+ \left( 1-\xi \right) R\ln  \left( 
\frac {20 \,{U}^{3/2}}{1-\xi} \right) 
+\xi\,R\ln  \left( \frac {5 \sqrt {2} \,{U}^{3/2}}{\xi} \right) \ ,
\label{feqig}
\end{equation}
where we used the variable $U=U_A+U_B = c (n_A+n_B) R T$ to rewrite the variables $U_A$ and $U_B$ as
\beq
U_A = \frac{ n_{A,0}-\xi}{n_{A,0}+n_{B,0}} U\ ,\quad
U_B = \frac{  n_{B,0}+\xi}{n_{A,0}+n_{B,0}} U \ .
\eeq
In the entropy representation $\Phi=S$, we choose the independent variables as 
$E^a=\{U,\xi\}$. Then, the metric of the equilibrium space (\ref{metrica}) for the chemical reaction of two ideal gases reduces to 
\beq
g_S^{ig} = \frac{dU^2}{U^2} 
+ \frac{  R d\xi^2} {{\xi}^{2}\left( 1-\xi \right) \left[ 1 - R \ln (2\sqrt{2}\, \xi) + R \ln(1-\xi) \right] }\ .
\label{gig}
\eeq
For the metric (\ref{gig})  the only non-vanishing Christoffel symbols are $\Gamma^U_{UU}$ and $\Gamma^\xi_{\xi\xi}$. Then, 
the geodesic equations read
\beq
{\frac {d^{2}}{d{\tau}^{2}}}U (\tau) - \frac{1}{U} 
\left( {\frac {d}{d\tau}}U (\tau)  \right) ^{2} = 0\ ,
\label{ec1} 
\eeq
\beq
{\frac {d^{2}}{d{\tau}^{2}}}\xi (\tau) 
+  \frac{ R - (2-3\xi)\left(1 - R \ln \frac{2\sqrt{2}\, \xi}{1-\xi}\right)}
                   {2\xi(1-\xi)\left(1 - R \ln \frac{2\sqrt{2}\, \xi}{1-\xi}\right)}
\left( {\frac {d}{d\tau}}\xi (\tau)  \right) ^{2} = 0 \ .
\label{ec2}
\eeq

The main point now is to see whether the geodesic equations can reproduce the thermodynamic process that occur during a 
chemical reaction. The idea is that we use the initial values of the thermodynamic variables,  corresponding to the initial equilibrium state of the chemical reaction, to identify a particular point in the equilibrium manifold. This point is then used as initial value to integrate the 
geodesic equations. The question is whether the solution of the geodesic equations passes through the final equilibrium state of the chemical reaction. 

\begin{figure}
        \centering
     \begin{subfigure}[b]{0.4\textwidth}
                \centering
               \includegraphics[scale=0.3]{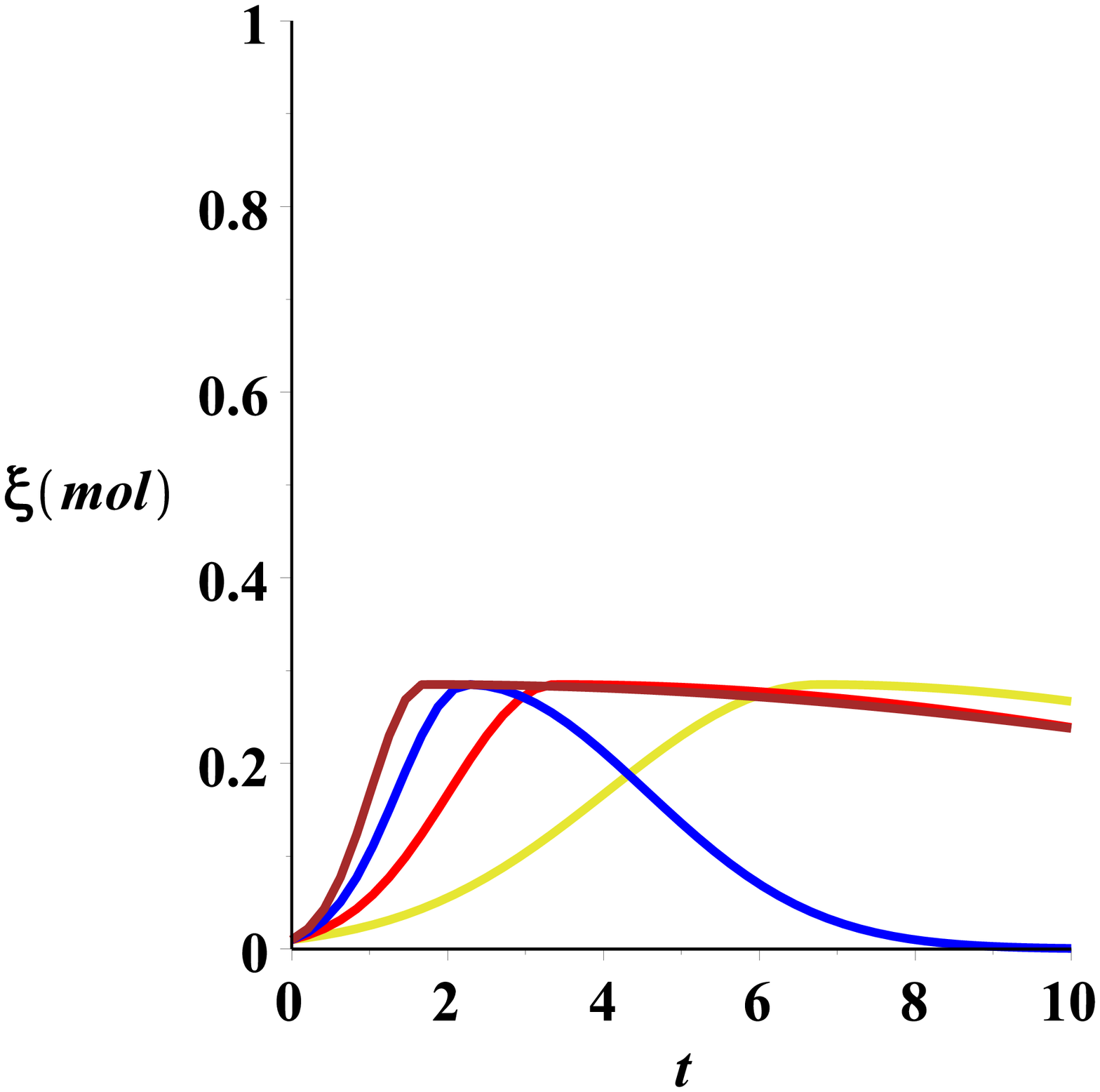}
\caption{Solution of geodesic equation \eqref{ec2} for $\xi(0)=0.01$ and different initial ``velocities'' $\dot{\xi}(0)$ }
\label{ideal1}
        \end{subfigure}
~
        \begin{subfigure}[b]{0.4\textwidth}
               \centering
\includegraphics[scale=0.3]{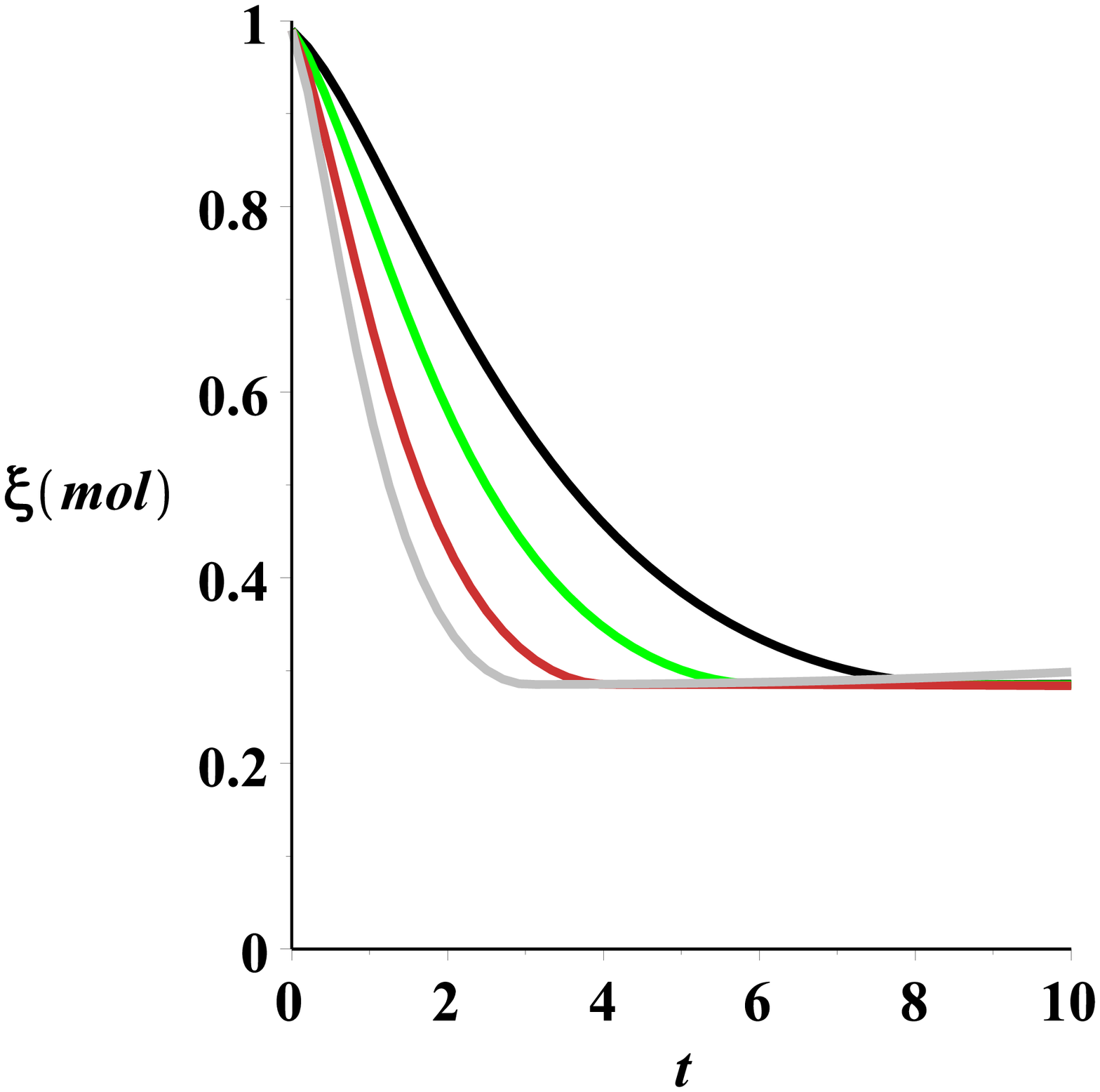}
\caption{Solution of geodesic equation \eqref{ec2} for $\xi(0)=0.99$ and different initial ``velocities'' $\dot{\xi}(0)$}
\label{ideal2}
        \end{subfigure}%
        ~ 
   
             \caption{Behaviour of the geodesic solution in the Entropy representation, for the reaction \ce{A_{(g)} <=> B_{(g)}} at T=300 K considering \ce{A} and \ce{B}  ideal gases }\label{fig:animals}
\end{figure}

For the 
particular case of ideal gases we are investigating here, we found in the last subsection that the thermodynamic analysis establishes the value
of the extent of reaction $\xi_f\approx 0.285$ for the final state. The values of the initial state have been incorporated in the 
fundamental equation (\ref{feqig}) and, consequently, in the thermodynamic metric (\ref{gig}) and in the geodesic equations. 
We now consider the ``experimental" condition $T = 300 K$. Then, we get $U(\tau) = \dfrac{3}{2} \, 8.314 \, 300 \, J=$ const.,
 so that equation \eqref{ec1} is satisfied identically.
Then,  we proceed to solve numerically the remaining geodesic equation \eqref{ec2} for $\xi$. 
The results are displayed in  figures \ref{ideal1} and \ref{ideal2}. We choose as initial conditions values very close to $\xi(0)= 0$ and $\xi(0)=1$ (since the reaction can go in both directions), and different arbitrary initial ``velocities'' $d \xi(0)/d \tau=\dot{\xi}(0)$. The important result is that all the geodesics reach the point $\xi_f\approx 0.285$, independently of the initial values of $\xi$ and $\dot\xi$, 
and all of them follow a pattern in the physical region between the initial value and $\xi_f$, where the entropy increases. In fact, the numerical 
integrator finds always a ``singularity" at the point  $\xi\approx 0.285$. We will see below that in fact this point corresponds to a coordinate
singularity of the thermodynamic metric.

We now test the Legendre invariance of our analysis by using the Massieu potential (\ref{massig}). We insert the conditions
for the present chemical reaction and obtain the corresponding fundamental equation. 
In this case, $E^a=\{\beta,\xi\}$ and the metric 
(\ref{metrica}) leads to 
\beq
g_\phi^{ig}= \frac{d\beta^2}{\beta^2} - \frac{R\, d\xi^2}{\xi^2(1-\xi)[1 + R \ln (2\sqrt{2}\, \xi) - R\ln (1-\xi)] }\ .
\eeq
The  only non-vanishing Christoffel symbols are $\Gamma^\beta_{\beta\beta}$ and $\Gamma^\xi_{\xi\xi}$. Then, the geodesic equations are 
\beq
{\frac {d^{2}}{d{\tau}^{2}}}\beta (\tau) - \frac{1}{\beta} 
\left( {\frac {d}{d\tau}}\beta (\tau)  \right) ^{2} = 0\ ,
\label{ec3} 
\eeq
\beq
{\frac {d^{2}}{d{\tau}^{2}}}\xi (\tau) 
-  \frac{ R + (2-3\xi)\left(1 + R \ln \frac{2\sqrt{2}\, \xi}{1-\xi}\right)}
                   {2\xi(1-\xi)\left(1 + R \ln \frac{2\sqrt{2}\, \xi}{1-\xi}\right)}
\left( {\frac {d}{d\tau}}\xi (\tau)  \right) ^{2} = 0 \ .
\label{ec4}
\eeq

\begin{figure}
        \centering
     \begin{subfigure}[b]{0.4\textwidth}
                \centering
               \includegraphics[scale=0.3]{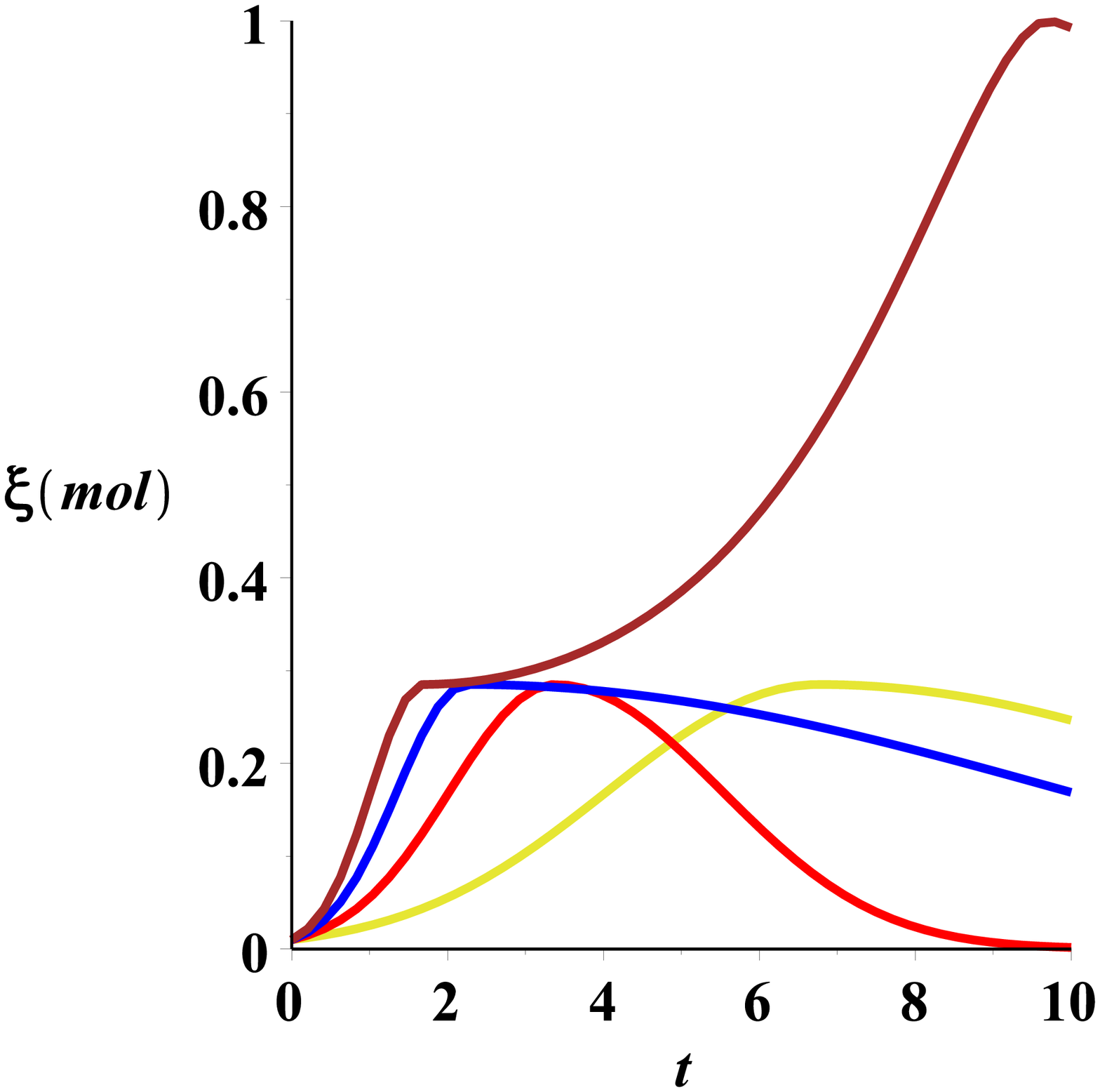}
\caption{Solution of geodesic equation \eqref{ec3} for $\xi(0)=0.01$ and different initial ``velocities'' $\dot{\xi}(0)$}
\label{ideal3}
        \end{subfigure}
~
        \begin{subfigure}[b]{0.4\textwidth}
               \centering
\includegraphics[scale=0.3]{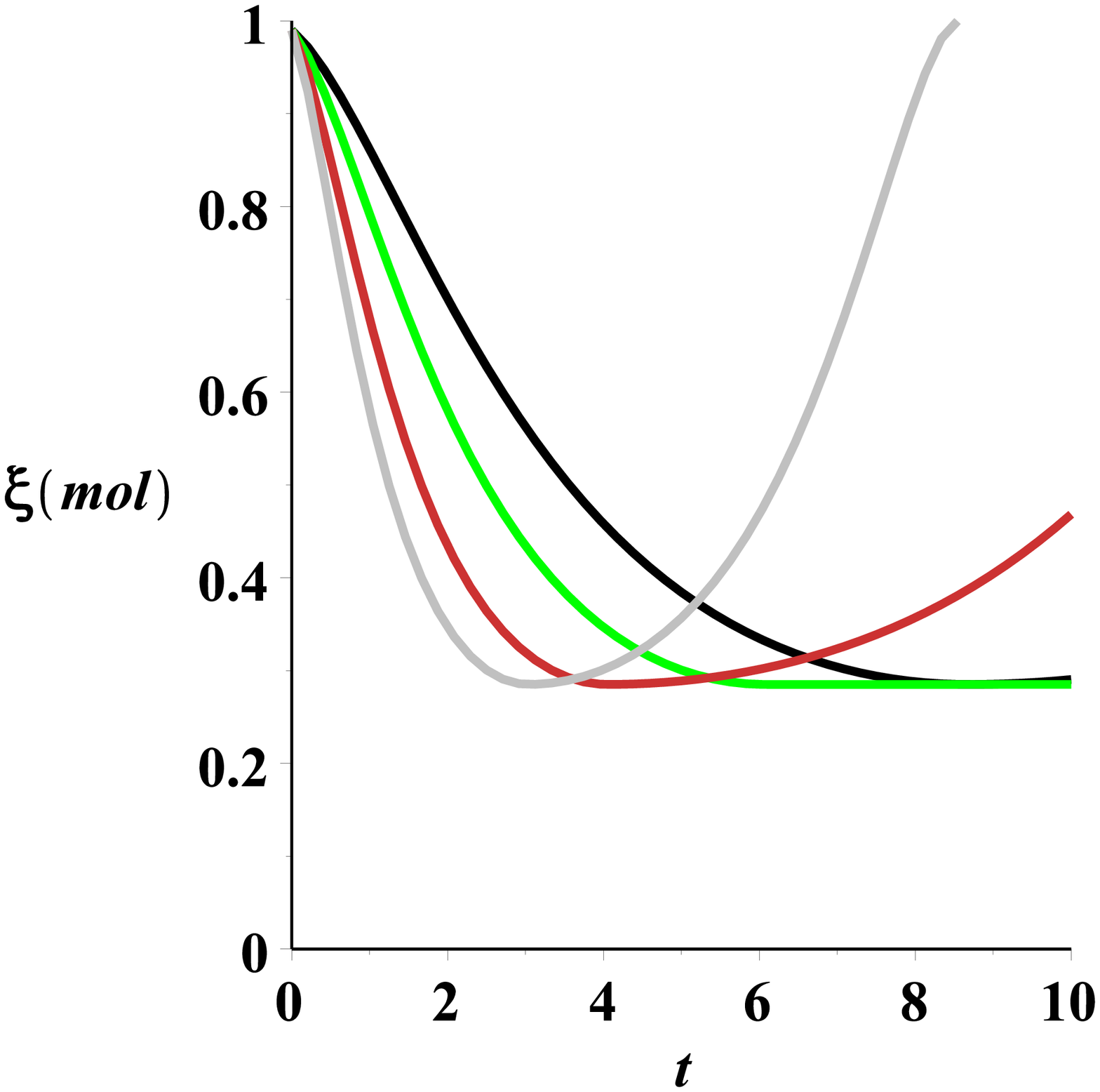}
\caption{Solution of geodesic equation \eqref{ec3} for $\xi(0)=0.99$ and different initial ``velocities'' $\dot{\xi}(0)$}
\label{ideal4}
        \end{subfigure}%
        ~ 
   
             \caption{Behaviour of the geodesic solution in the Massieu Potential representation, for the reaction \ce{A_{(g)} <=> B_{(g)}} at T=300 K considering \ce{A} and \ce{B}  ideal gases }\label{fig:animals1}
\end{figure}

We now fix the ``experimental'' condition $\beta(\tau) = \dfrac{1}{300} K^{-1}$, with that the equation \eqref{ec3} is satisfied identically and we  proceed to solve numerically the  remaining 
equation (\ref{ec4}) for $\xi$. The initial  conditions are the same as in the entropy representation. The results are displayed in figures \ref{ideal3} and \ref{ideal4}.
We see again that all the geodesics pass through the point $\xi_f\approx 0.285$, independently of the initial values. Moreover, in the physical region, before the final equilibrium state is reached, they all coincide with the geodesics shown in Figs. \ref{ideal1} and \ref{ideal2} for the analysis in the entropy representation. This shows explicitly that the analysis does not depend on the choice of thermodynamic potential.

A straightforward computation shows that the curvature tensor vanishes in both representations and, consequently, the corresponding equilibrium space is flat.  
This indicates that no thermodynamic interaction exists in a chemical reaction in which only non-interacting ideal gases are involved.

\section{Van der Waals gases}
\label{sec:vdw}

\subsection{Thermodynamics}

A more realistic gas is described by the van der Waals fundamental equation  \cite{Callen}
\beq
S(U,V,n) = n s_0 +  nR \ln\left[\left(\frac{{\frac {U}
{n}}+{\frac {an}{V}} }{cRT_0}\right)^c \frac{n_0}{V_0} \left( \frac{V}{n}-b\right)\right]\ ,
\eeq
where $a$ and $b$ are constants. 
For simplicity, we will consider both gases with the same $a$ value, so that the coupling terms in the fundamental equation \eqref{total} 
can be considered as vanishing. That is, this case is a simple  mixture in which the interactions $A-B$ are identical to the interactions $A-A$ and $B-B$. 
Consequently, the fundamental equation  of the mixture reads
\beq
\label{mas}
S(U_A,U_B,V,n_A,n_B) = S_A(U_A,V,n_A) + S_B(U_B,V,n_B) \ .
\eeq

In the previous section, the constant temperature condition of the ideal gas reaction was equivalent to implying $U_{total}=$const. and, therefore, 
it was straightforward to compare the $S$ and $\phi$ representations. In fact, the conditions, under which the extremum principle is valid, were fulfilled 
in both representations.  On the other hand, in the case of van der Waals gases, the constant temperature condition does not imply that the total 
internal energy is constant. Consequently, it is necessary to be cautious when applying the extremum principle corresponding
to a constant temperature condition. To this end, we perform a Legendre 
transformation in the $S-$representation to obtain the Massieu potential $\phi$ so that the 
temperature is an independent variable.  Then, for the chemical reaction of two van der Waals gases we obtain the 
thermodynamic potential
\begin{eqnarray}
\phi(\beta,V,n_A,n_B) &=& \phi_A(\beta,V,n_A) + \phi_B(\beta,V,n_B) \notag \\
&=&-\beta\left[ \mu_A(\beta,V,n_A)n_A - p_AV + \mu_B(\beta,V,n_B)n_B - p_BV\right] \ .
\label{feqvdw}
\end{eqnarray}

\begin{figure}
\centering
 \includegraphics[scale=0.3]{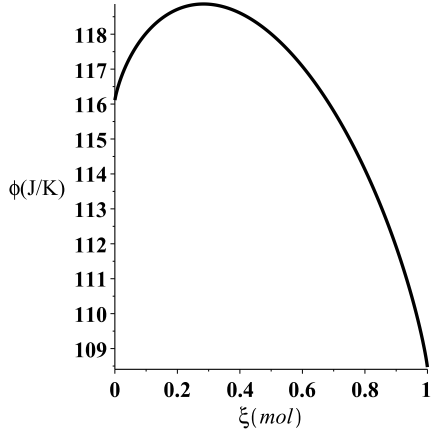}
\caption{Behavior of the thermodynamic potentials for the reaction \ce{A_{(g)} <=> B_{(g)}} at $T=300 K$ considering \ce{A} and \ce{B}  
as van der Waals gases.}
\label{vdwmas}
\end{figure}

For the sake of simplicity, we consider the same initial values as given in Table \ref{condiciones} for ideal gases, with the additional values
$a=506\,J\, L\,mol^{-2}$ and $b=0.050\, L \, mol^{-1}$, which are quite reasonable for real gases \cite{CRC}. 
The behavior of the potential is displayed in Fig. \ref{vdwmas}. 
Notice that in this case the maximum is reached at $\xi=\xi_f\approx 0.284$ which is only slightly different from the ideal case,
because the initial equilibrium state is the same in both cases and, at $T=300$ K, the differences between ideal and van der Waals gases 
are very small. 
 In the next section, we will find out if there exists a geodesic in the equilibrium space that connects the initial equilibrium state with the final one.

\subsection{Geometrothermodynamics}

Considering the initial conditions given in Table \ref{condiciones}, the fundamental equation in this case can be obtained 
directly from Eq.(\ref{feqvdw}) as
\begin{eqnarray}
\phi(\beta,\xi) =& &1 - cR\left(1+\ln\frac{\beta}{cR}\right) + \xi \left(1-cR\ln 2 + R \ln\frac{20-b\xi}{\xi}\right)\nonumber\\
& &-(1-\xi)\left(\frac{1}{10} a\beta \xi -R\ln\frac{20-b+b\xi}{1-\xi}\right)\ .
\end{eqnarray}
From here one can compute all the components of the metric tensor. The explicit expressions are quite cumbersome and cannot be written in 
a compact form. The curvature scalar is in general different from zero, indicating the presence of thermodynamic interaction. This is 
in accordance with the statistical approach to the van der Waals gas in which, as a result of the interaction between the particles of the gas,
 the corresponding Hamiltonian possesses a non-trivial potential term.

\begin{figure}
        \centering
     \begin{subfigure}[b]{0.4\textwidth}
                \centering
               \includegraphics[scale=0.3]{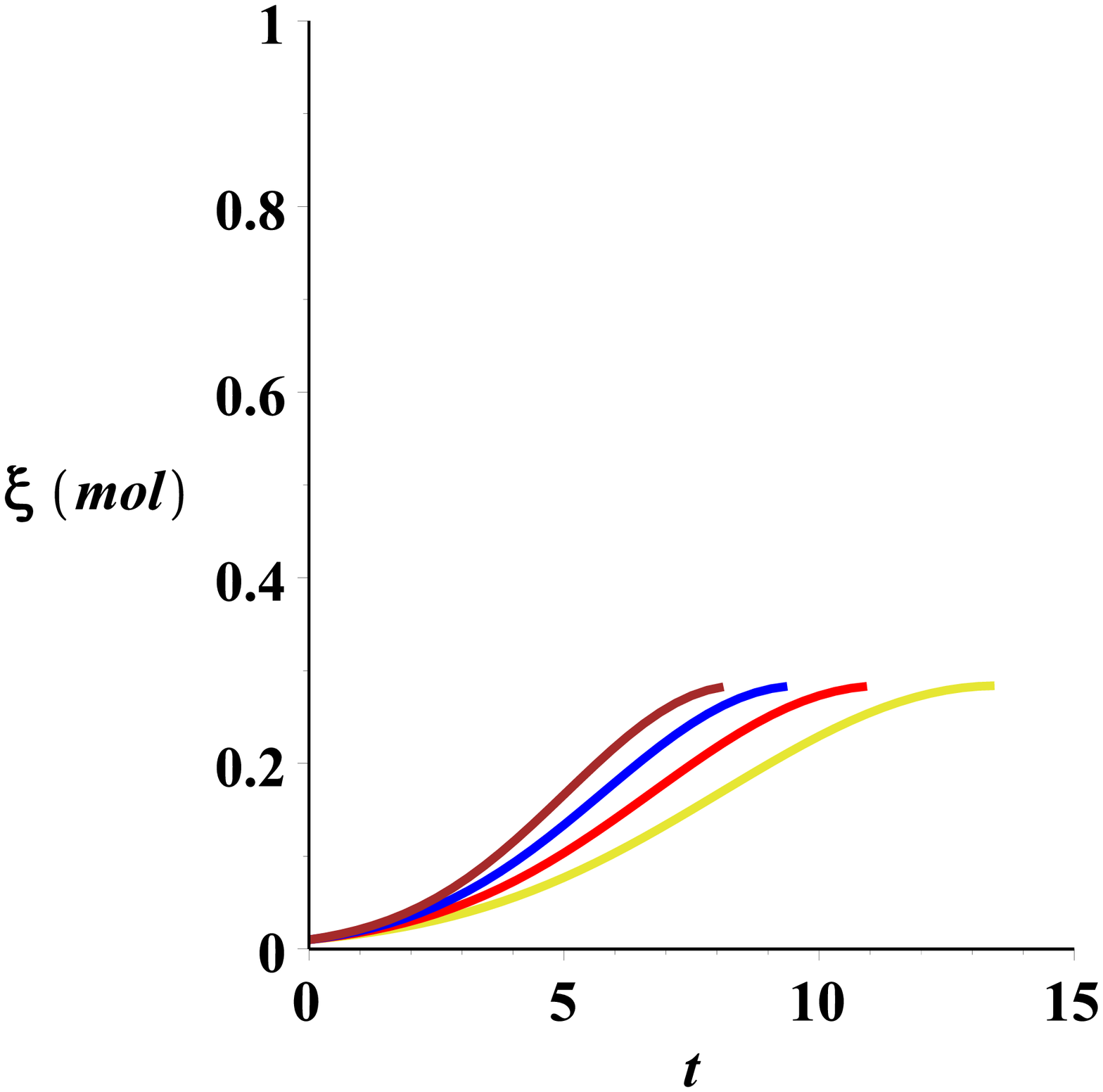}
\caption{Solution of geodesic equation \eqref{ec5} for $\xi(0)=0.01$ and different initial ``velocities'' $\dot{\xi}(0)$}
\label{vdw1}
        \end{subfigure}
~
        \begin{subfigure}[b]{0.4\textwidth}
               \centering
\includegraphics[scale=0.3]{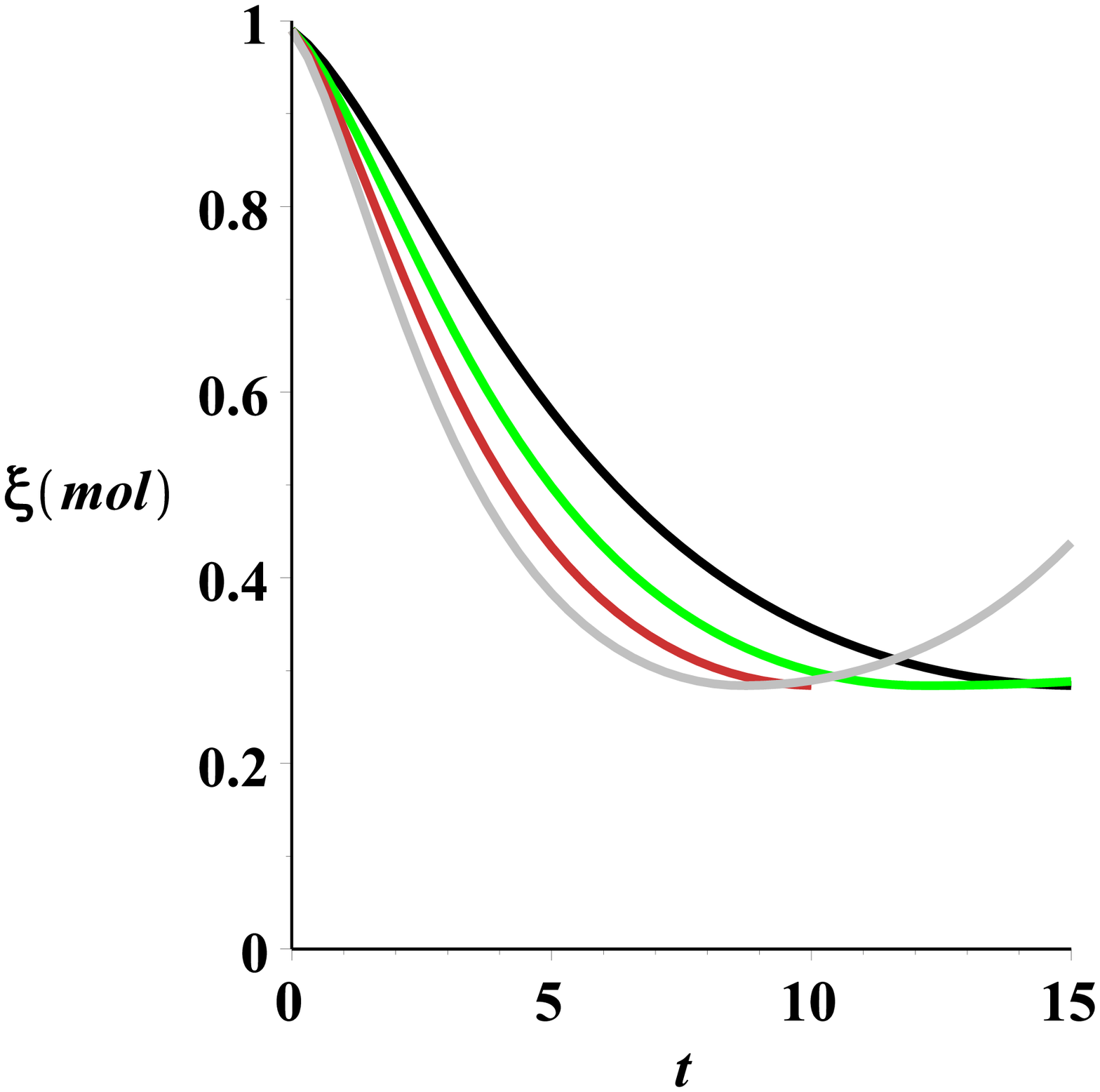}
\caption{Solution of geodesic equation \eqref{ec5} for $\xi(0)=0.01$ and different initial ``velocities'' $\dot{\xi}(0)$}
\label{vdw2}
        \end{subfigure}%
        ~ 
   
             \caption{Behaviour of the geodesic solution in the Massieu Potential representation, for the reaction \ce{A_{(g)} <=> B_{(g)}} at T=300 K considering \ce{A} and \ce{B} Van der Waals gases }\label{geodvdw}
\end{figure}

All the Christoffel symbols are different from zero so that the geodesic equations contain all the independent terms 
\begin{eqnarray}
{\frac {d^{2}}{d{\tau}^{2}}}\beta (\tau) + \Gamma^\beta_{\beta\beta} \left( {
\frac {d}{d\tau}}\beta (\tau)  \right) ^{2}+ 2\Gamma^\beta_{\beta\xi}\left(\frac {d}{d\tau} \beta (\tau)\right)\left( \frac {d}{d\tau}\xi (\tau)\right) + \Gamma^\beta_{\xi\xi} \left( {
\frac {d}{d\tau}}\xi (\tau)  \right) ^{2}= 0 \ ,\\
{\frac {d^{2}}{d{\tau}^{2}}}\xi (\tau) + \Gamma^\xi_{\xi\xi} \left( {
\frac {d}{d\tau}}\xi (\tau)  \right) ^{2}+ 2\Gamma^\xi_{\beta\xi}\left(\frac {d}{d\tau} \beta (\tau)\right)\left( \frac {d}{d\tau}\xi (\tau)\right)+  \Gamma^\xi_{\beta\beta} \left( {
\frac {d}{d\tau}}\beta (\tau)  \right) ^{2} = 0 \ .
\label{ec5}
\end{eqnarray}
This system is solved numerically with initial values for $\xi$ very close to $\xi(0)= 0$ and $\xi(0)=1$ and arbitrary initial  ``velocities'' $\dot{\xi}(0)$. Moreover, 
for $\beta$ we introduce the ``experimental'' condition in the form $\beta(0) = \dfrac{1}{300}$ and $\dot{\beta(0)} = 0$.
The results are displayed in Figs. \ref{vdw1} and \ref{vdw2}. We see also in this case that all the geodesics reach their 
final equilibrium state for $\xi=\xi_f\approx 0.284$ which is the same value we obtained in the thermodynamic analysis of the 
last subsection. Moreover, the geodesics follow a pattern  in the physical region, in the same way as in the case of ideal gases. 
We conclude that the geodesics can be interpreted also in this case as the geometric path that the chemical reaction follows until it reaches its final equilibrium state. 

\section{Geometrothermodynamics of chemical reactions}
\label{sec:gtdcr}

In the previous sections, we analyzed two particular chemical reactions with only two species. In both cases, we 
found that the corresponding equilibrium manifold ${\cal E}$ reduces to a $2-$dimensional manifold, once the conditions of the reaction
are taken into account. The question arises whether the dimension of ${\cal E}$ increases as the number of reactants and products of the 
reaction increases. We will show in this section that GTD can handle in a simple manner the chemical reaction of 
any arbitrary (finite) number of species in a closed system.   

Consider the chemical reaction of $r$ species described by the variables  $S_i$, $U_i$, $V_i$, and $n_i$ $(i=1,...,r)$. In the entropy representation, for instance, the thermodynamic properties of each substance is determined by the fundamental 
equation $S_i=S_i(U_i,V_i,n_i)$. Each fundamental equation $S_i$ generates a $3-$dimensional equilibrium space ${\cal E}_i$ for the $i-$species. According to Eq.(\ref{total}), the fundamental equation of the chemical system $\Phi=\Phi(S_1,...,S_r)= \Phi(U_i,V_i,n_i)$ will depend on $3r$ variables. Then, the total equilibrium space ${\cal E}$ has $3r$ dimensions. In general, the level of computational difficulty in geometry increases with the number of dimensions, so that for large $r$ the calculations could easily be outside the reach of computational capability. 
However, we can use the conditions of the reaction to reduce the number of dimensions. A common condition for reactions involving gases is that the reaction occurs at 
constant volume so that the functional dependence of the fundamental equation can be reduce to $2r$, i.e.,  $\Phi=\Phi(U_i,n_i)$. 
Furthermore, using the definition of the extent of reaction parameter, $\Delta n_i = \nu_i \xi$, we can replace  all the $n_i$'s variables
by $\xi$, according to $n_i = n_{i,0} + \nu_i \xi$. Since $n_{i,0}$ and $\nu_i$ are constants, the functional dependence of 
the fundamental equation reduces to $\Phi=\Phi(U_i,\xi)$.  Using the equations of state for each species, we can express each $U_i$ in terms of $U$, $n_i$ and other constants. For instance, in the case of ideal gases we have that 
\beq
U= \sum_{i} U_i = \sum_i c_i n_i R T = RT \sum_i c_i (n_{i,0} + \nu_i \xi) \ .
\eeq
Then, we can express each $U_i$ as 
\beq
U_i(U,\xi) = \frac{c_i (n_{i,0} + \nu_i\xi)}{\sum_j c_j (n_{j,0}+\nu_j \xi)} U \ ,
\eeq
so that the fundamental equation becomes $\Phi=\Phi(U,\xi)$. Consequently, the corresponding equilibrium manifold is 2-dimensional,
independently of the number of species. 

In the case of more complicated fundamental equations, it is always possible to express each $U_i$ in terms ofU and the 
extent of reaction in such a way that the resulting equilibrium manifold has only two dimensions. For instance, in the case of van der Waals gases, we obtain
\beq
U_i(U,\xi) = \frac{c_i(n_{i,0}+\nu_i \xi)}{\sum_j c_j (n_{j,0}+\nu_j \xi)}\left(U + \frac{a}{V} \sum_j (n_{j,0}+\nu_j \xi)^2\right) -
\frac{a}{V} (n_{i,0}+\nu_i \xi)^2 \ .
\label{redvdw}
\eeq

Let us now investigate the geodesic equations. 
To construct the metric $g$ of the equilibrium manifold ${\cal E}$, we can use the thermodynamic potential $\Phi(U,\xi)$ or any other potential that can be obtained from  $\Phi(U,\xi)$ by means of a Legendre transformation (basically, $\tilde \Phi(\beta,\xi)$). The results do not depend on the choice of $\Phi$,  because the geometric properties of ${\cal E}$ in GTD are Legendre invariant. If we take the potential $\Phi(U,\xi)$, the geodesic equations can be written in general as
\begin{eqnarray}
{\frac {d^{2}}{d{\tau}^{2}}}U  (\tau) + \Gamma^U _{U U } \left( {
\frac {d}{d\tau}}U  (\tau)  \right) ^{2}+ 2\Gamma^U _{U \xi}\left(\frac {d}{d\tau} U  (\tau)\right)\left( \frac {d}{d\tau}\xi (\tau)\right) + \Gamma^U _{\xi\xi} \left( {
\frac {d}{d\tau}}\xi (\tau)  \right) ^{2}= 0 \ ,\\
{\frac {d^{2}}{d{\tau}^{2}}}\xi (\tau) + \Gamma^\xi_{\xi\xi} \left( {
\frac {d}{d\tau}}\xi (\tau)  \right) ^{2}+ 2\Gamma^\xi_{U \xi}\left(\frac {d}{d\tau} U  (\tau)\right)\left( \frac {d}{d\tau}\xi (\tau)\right)+  \Gamma^\xi_{U U } \left( {
\frac {d}{d\tau}}U  (\tau)  \right) ^{2} = 0 \ .
\label{ec6}
\end{eqnarray}

According to the Picard-Lindel\"of theorem \cite{plt}, given an initial value, 
i.e.,  $\xi(0)$ , $\dot \xi(0)$, $U(0)$ and $\dot U(0)$,
if ${\cal E}$ is smooth -as in the preceding cases-, the solution to this equation exists and is unique. 
In the case of 
a chemical reaction,  the  value of $\xi(0)$ is fixed by the initial equilibrium state of the reaction, but the value of $\dot\xi(0)$ remains 
free. In the case of ideal gases, in which $U$ is proportional to $T$, the initial values of the second variable $U(0)$ and $\dot U(0)$ are fixed by the conditions of the chemical reaction. In the case of more general species, a Legendre transformation must be performed such that $T$ becomes the second independent variable and the initial values are $T(0)$ and $\dot T(0)$. 

We have seen in the examples above that the final equilibrium state of the reaction does not depend on $\dot\xi(0)$ and that the numerical integrator detects a ``singularity" at that point. It turns out that this corresponds to a coordinate singularity of the metric. To see this, we calculate the component $(\xi\xi)$ of the general metric \eqref{metrica}, and obtain 
\beq
\label{componente}
g_{\xi\xi} =  \left(\xi \frac{\partial \Phi}{\partial \xi}\right) ^{-1} \frac{\partial^2 \Phi}{\partial \xi^2}  
= \left(\xi \sum_i \frac{\nu_i \,\mu_i}{T}\right)^{-1} \left(\sum_i \frac{\nu_i}{T} \right) \frac{\partial \mu_i}{\partial \xi}\ .
\eeq
This expression is valid for $\Phi=S$ and all the thermodynamic potentials  that can be obtained from $S$ by means of any 
Legendre transformations -except those which change the role of $\xi$-. 
It is easily seen that as soon as the chemical-reaction 
equilibrium condition \eqref{equilib} is satisfied, the denominator goes to zero and, thus, the metric is not well defined. 
This means that at the final equilibrium point the coordinates are not appropriate to describe the equilibrium manifold. Notice
that this result is completely general since it does not depend on the particular reaction that determines the potential $\Phi$.

The fact that the final equilibrium state is characterized by a coordinate singularity  allows us to perform an
analytical investigation of that particular point. Consider, for instance, the metric (\ref{gig}) for two ideal gases
in the $S-$representation. The component $g_{\xi\xi}$ presents a physical divergence when  
$1 - R \ln (2\sqrt{2}\, \xi) + R \ln(1-\xi)=0$. The solution of this equation
\beq
\xi_f= \frac{1}{1+2\sqrt{2}\, e^{-1/R}}
\eeq
with $R=8.314$ gives $\xi_f \approx 0.285$ which is exactly the value obtained in the numerical investigating of the geodesic 
equations for this metric.

Finally, let us mention an additional invariance property of the GTD approach. 
To reduce the number of independent variables of the fundamental equation in the case of van der Waals gases, we used the relationship (\ref{redvdw}) for $U_i(U,\xi)$ with $V=const$. However, due to the invariance of the metric $g$ under changes of coordinates, we can also use the same relationship as $U_i=U_i(V,\xi)$. Then, the resulting fundamental equation becomes $\Phi(V,\xi)$, once $\beta$ is fixed in accordance 
with the conditions of the reaction.   
The corresponding metric can be computed and the geodesic equations can be integrated
numerically for $V$ and $\xi$ with the same initial conditions for $\xi$, and $V(0) = 20$ and $\dot V(0) = 0$. 
The resulting geodesics are exactly the same as the ones we obtained with $\Phi(U,\xi)$.

\section{Conclusions}
\label{sec:con}

In this work, we used the formalism of GTD to present a geometric representation of chemical reactions in closed systems.
In GTD, all the information about a thermodynamic system is encoded in its equilibrium manifold determined by 
a metric which is invariant under Legendre transformations, i.e, its properties do not depend on the choice 
of thermodynamic potential.

First, we consider the case of a chemical reaction with only two species corresponding either to ideal gases or to van der Waals gases. 
In the case of ideal gases, we found that the equilibrium manifold is flat, independently of the thermodynamic potential.  
In GTD, a vanishing curvature means that there is no thermodynamic interaction. This agrees with the interpretation from the point of view of  statistical mechanics and thermodynamics: With the statistical point of view, because the molecules of each ideal gas behave as ``free particles" since the Hamiltonian contains exclusively the kinetic part; and with the thermodynamic point of view,  
because there are no  coupling terms in the fundamental equation \eqref{fundamental}. 
In the case of van der Waals gases, the curvature is different from zero, indicating the presence of thermodynamic interaction. This is also 
in accordance with the statistical approach to the van der Waals system, because the Hamiltonian contains a potential term which is responsible
for the interaction between the molecules of the system.

The thermodynamic analysis of the chemical reactions in both cases shows that the final equilibrium state is reached at a particular value of the extent of reaction parameter. The numerical analysis of the geodesics in the corresponding equilibrium manifolds provides exactly the same value of the extent of reaction for the final equilibrium state. It is evident that all the geodesics follow a path pattern that ends at 
the same equilibrium state, independently of the initial values of $\xi(0)$ and $\dot{\xi}(0)$. Moreover, the final equilibrium state is 
always denoted as a singularity by the numerical integrator. 

In order to understand our results in a more general fashion, we analyzed a general reaction in the context of GTD. We showed that using 
the conditions of thermodynamic equilibrium and the laboratory conditions of the reaction, it is always possible to reduce to two the 
number of dimensions of the equilibrium manifold. This is an interesting result that allows us to describe any chemical reaction as 
a geodesic curve on a $2-$dimensional space. Finally, it was shown that the metric of the equilibrium manifold possesses a coordinate 
singularity exactly at that point where the condition for the reaction equilibrium is satisfied. 

The examples of chemical reactions presented in this work involve only gases. Nevertheless, the generalization to include reactions involving  
solids or liquids is straightforward. Indeed, once the fundamental equations of the species are given, we can construct the corresponding 
Massieu-Planck potential of the reaction which allows us to consider $P$ and $T$ as constants.

The main conclusions of this work is that to any chemical reaction in a closed system we can associate a $2-$dimensional 
equilibrium manifold, and that any chemical reaction can be represented as a geodesic in which the initial state is determined by the initial 
conditions of the reaction and the final state corresponds to a coordinate singularity of the thermodynamic metric.

\section{Acknowledgments}

We would like to thank the members of the GTD-group at the UNAM for fruitful comments and discussions.
This work was supported by CONACyT-Mexico, Grant No. 166391,  and by CNPq-Brazil. One of us (DT) would like
to thank CONACyT, CVU No. 442828, for financial  support.


\begin{thebibliography}{99}



\bibitem{gibbs}
J. Gibbs, \textit{The collected works}, Vol. 1, Thermodynamics (Yale University Press, 1948).

\bibitem{cara}
C. Charatheodory, Untersuchungen uber die Grundlagen der Thermodynamik, Gesammelte
 ̈ Mathematische Werke, Band 2 (Munich, 1995).



\bibitem{her73} R. Hermann, {\it Geometry, physics and systems} (Marcel
Dekker, new York, 1973).


\bibitem{mrugala1}
R. Mrugala, {\it Geometrical formulation of equilibrium phenomenological thermodynamics}, Rep.
Math. Phys. 14, 419 (1978).

\bibitem{mrugala2}
R. Mrugala, {\it Submanifolds in the thermodynamic phase space}, Rep. Math. Phys. 21, 197 (1985).



\bibitem{rao45} C. R. Rao, 
{\it Information and the accuracy attainable in the estimation of statistical parameters},
Bull. Calcutta Math. Soc. {\bf 37}, 81 (1945).

\bibitem{amari85} S. Amari, {\it Differential-Geometrical Methods in Statistics} (Springer-Verlag, Berlin, 1985).

\bibitem{quev07} H. Quevedo, {\it Geometrothermodynamics}
J. Math. Phys.
{\bf 48}, 013506 (2007).

\bibitem{qr12} H. Quevedo and A. Ramirez, {\it A geometric approach to the thermodynamics of the van der Waals system}
in Proceedings of Mario Novello's 70th Anniversary Symposium,
edited by N. Pinto-Neto and S. E. Perez-Bergliaffa (2012) pp. 301-313;
arXiv:1205.3544

\bibitem{qstv10a} H. Quevedo, A. S\'anchez, S. Taj, and A. V\'azquez, 
{\it Phase transitions in geometrothermodynamics}
Gen. Rel. Grav. {\bf 43}, 1153 (2011).

\bibitem{abcq12} A. Aviles, A. Bastarrachea, L. Campuzano and H. Quevedo, {\it 
Extending the generalized Chaplygin gas model by using geometrothermodynamics}
Phys. Rev. D {\bf 86}, 063508 (2012).
 
 
\bibitem{arnold} V. I. Arnold, {\it Mathematical Methods of Classical Mechanics}
(Springer Verlag, new York, 1980).

\bibitem{vqs10} A. V\'azquez, H. Quevedo, and A. S\'anchez, 
{\it Thermodynamic systems as extremal hypersurfaces}
J. Geom. Phys. {\bf 60}, 1942 (2010).


\bibitem{blnq13} A. Bravetti, C. Lopez, F. Nettel and H. Quevedo, {\it Change of representations in geometrothermodynamics}
(2013) in preparation.


\bibitem{Levine}
I. N. Levine, (2009), {\it Physical Chemistry} (McGraw-Hill, New York, USA, 2009).



\bibitem{Callen}
H. B. Callen, {\it  Thermodynamics and an Introduction to Thermostatics}
(John Wiley and Sons, Inc., New York, 1985).


\bibitem{Tschoegl}
N. W. Tschoegl, {\it Fundamentals of Equilibrium and Steady-State thermodynamics} (Elsevier Publishers, 
Amsterdam, The Netherlands, 2000).


\bibitem{CRC}
D.R. Lide (ed.), (2010), {\it CRC Handbook of Chemistry and Physics}, (CRC Press, London, UK, 2010).


\bibitem{plt} Y, Choquet-Bruhat, C. Dewitt-Morette, and M. Dillard-Bleick, 
{\it Analysis, manifolds and physics} 
(Elsevier Publishers, Amsterdam, The Netherlands, 1982).

\end{thebibliography}
\end{document}